# Estimating the spectrum in computed tomography via Kullback-Leibler divergence constrained optimization

Wooseok Ha, Emil Y. Sidky, Rina Foygel Barber, Taly Gilat Schmidt, and Xiaochuan Pan

04.30.18

## Abstract

**Purpose:** We study the problem of spectrum estimation from transmission data of a known phantom. The goal is to reconstruct an x-ray spectrum that can accurately model the x-ray transmission curves and reflects a realistic shape of the typical energy spectra of the CT system.

**Methods:** Spectrum estimation is posed as an optimization problem with x-ray spectrum as unknown variables, and a Kullback-Leibler (KL) divergence constraint is employed to incorporate prior knowledge of the spectrum and enhance numerical stability of the estimation process. The formulated constrained optimization problem is convex and can be solved efficiently by use of the exponentiated-gradient (EG) algorithm. We demonstrate the effectiveness of the proposed approach on the simulated and experimental data. The comparison to the expectation-maximization (EM) method is also discussed.

**Results:** In simulations, the proposed algorithm is seen to yield x-ray spectra that closely match the ground truth and represent the attenuation process of x-ray photons in materials, both included and not included in the estimation process. In experiments, the calculated transmission curve is in good agreement with the measured transmission curve, and the estimated spectra exhibits physically realistic looking shapes. The results further show the comparable performance between the proposed optimization-based approach and EM.

**Conclusions:** Our formulation of a constrained optimization provides an interpretable and flexible framework for spectrum estimation. Moreover, a KL-divergence constraint can include a prior spectrum and appears to capture important features of x-ray spectrum, allowing accurate and robust estimation of x-ray spectrum in CT imaging.

# 1 Introduction

In x-ray imaging, determination of the x-ray energy spectrum is an important task for many applications, including patient dose calculation, beam-hardening correction, and dual-energy material decomposition (DeMarco et al. [1], Heismann et al. [2], Engler and Friedman [3]). For an energy resolved CT system, the x-ray spectrum represents the product of the polychromatic source spectrum and the detector spectral response. Interest in estimating the x-ray spectrum of a CT system is recently growing due to the development of spectral CT with a photon-counting detector (Taguchi and Iwanczyk [4], Schlomka et al. [5], Schmidt et al. [6]).

One common approach for reconstructing the x-ray energy spectrum is based on transmission measurements acquired by a CT scanner through a calibration phantom of known thicknesses and materials. This method formulates the measurement process of x-ray photons into a linear system of equations, and after acquiring calibration transmission measurements, the x-ray spectrum is recovered by inverting the linear system. While this inverse problem of spectrum estimation is intrinsically unstable due to its ill-conditioning, a number of methods have been proposed for obtaining a stable and accurate solution.

Using a physical model with few parameters effectively reduces the degrees of freedom of the problem, and allows for stable estimation of the spectrum by expressing a low-dimensional representation of the x-ray spectrum (Silberstein [7], Perkhounkov et al. [8], Zhao et al. [9]). The parameters are fitted with least squares or other data discrepancy objectives. Meanwhile, another line of work investigates an iterative perturbation method (e.g. Waggener et al. [10]) that minimizes differences between measured and calculated transmission curves using low-Z attenuators.

Various forms of regularization have been also employed to avoid the ill-conditioning of the problem and ensure stable spectrum estimation. For instance, a minimization of the sum of a $\chi^2$ objective term and a nonlinear regularization term is performed in Ruth and Joseph [11] to stabilize the final solution. The expectation-maximization (EM) method (Sidky et al. [12]) iteratively solves the ill-conditioned linear system and truncates the iteration of the algorithm at some finite iteration. Here early stopping serves as a sort of regularization as it prevents overfitting of the model. Singular value decomposition (SVD) is a more direct approach that attempts to directly invert the linear system to estimate each bin contents of spectrum (Francois et al. [13], Stampanoni et al. [14]). The SVD method often involves truncating smaller singular values and singular vectors of the system matrix, also known as truncated singular value decomposition (TSVD), since these components make almost no contribution to the measured data and are susceptible to the noise (Armbruster et al. [15]). The obtained spectrum from TSVD is sufficiently accurate to model the measured transmission curve, but it has the drawback that positivity of the spectrum is not guaranteed and the solution exhibits negative values in some energy positions. Recently, an extension of the TSVD method, called prior truncated singular value decomposition (PTSVD), has been proposed in Leinweber et al. [16] to further incorporate prior information about the statistical nature of the transmission data and about high-frequency spectral components such as characteristic peaks. In particular, by exploiting basis vectors for the null space of the system matrix, the authors reconstruct an x-ray spectrum that accurately reproduces the physical shape of the ground truth spectrum.

In this work, we present a new x-ray spectrum reconstruction method based on transmission measurements of a calibration phantom. Our aim is to formulate spectrum estimation as an optimization problem, for which an efficient first-order iterative algorithm is employed to solve the resulting optimization problem rapidly. The proposed method is capable of incorporating prior information about the physical shape of an x-ray spectrum, which enables accurate and realistic estimation of x-ray spectrum by including the characteristic lines of the target spectrum in the final estimation. Although the method can be used for any spectrum estimation task, in this work the focus is on photon-counting CT. The effective spectrum estimate, which includes the source spectrum and detector response, is needed for material decomposition into basis material sinograms (Schlomka et al. [5]) and for direct inversion into basis material images (Schmidt et al. [6], Barber et al. [17]). Our optimization-based approach can also be useful when the spectral calibration of an imaging system is combined with other optimization-based algorithms for spectral CT image reconstruction. A simulation study is carried out to demonstrate the utility of the method on spectrum determination, and the method is further evaluated on the measured transmission data through a step wedge phantom.

## 2 Materials and methods

### 2.1 Transmission measurement model

We assume the standard transmission measurement model for an x-ray imaging system—writing $\hat{c}_\ell$ to denote the number of transmitted photon counts along ray $\ell$ which encodes different source positions, then the forward data model after discretization is expressed as

$$\hat{c}_\ell = N_\ell \sum_i s_i \exp\left\{-\sum_m x_{m\ell}\mu_{mi}\right\}, \tag{1}$$

where $N_\ell$ is the expected number of photon counts detected along ray $\ell$ in the absence of an object, $s_i$ is the normalized distribution of x-ray photons at frequency $i$ in the absence of an object, $\mu_{mi}$ is the linear x-ray attenuation coefficient for material $m$ at frequency $i$, and $x_{m\ell}$ is the total amount of material $m$ lying along ray $\ell$. The model given in (1) is idealized and neglects numerous physical factors such as x-ray scatter.

The x-ray spectrum, given by $s_i$ across frequencies indexed by $i$, comprises the energy spectrum of the x-ray source and the spectral response of the detector. While we may assume that the x-ray source spectrum can be modeled to a certain degree, the detector spectral response is typically unknown due to many non-ideal physical effects of the detector. For instance, photon-counting detectors can discriminate incident x-ray photons based on their energies, allowing for the CT data acquisition in each energy window, and thus are useful for material decompositions to more than two basis functions; however, they also exhibit undesirable technical issues such as pulse pile-up and charge sharing (Taguchi and Iwanczyk [4]), potentially resulting in serious artifacts in the reconstructed images. Therefore, in reconstructing photon-counting CT images, it is crucial to accurately calibrate the spectral response of the detectors for further reduction of image artifacts.

The spectrum estimation approach studied in this work inverts the forward model (1) to estimate the x-ray spectrum, $\{s_i\}$, from noisy transmission measurements, $\{c_\ell\}$. The difficulty inherenet in inverting (1), however, is twofold.

First, the system matrix describing the attenuation of x-ray photons is highly low-rank, leading to the ill-conditioned linear system of the spectrum estimation problem. In particular, some form of regularization is necessary for reliable estimation of the x-ray spectrum. Second, the physical nature of an x-ray spectrum involves multiple structures in its shape, namely, the low-frequency component arising from bremsstrahlung radiation, which covers the entire range of the energy bins, and the high-frequency component arising from characteristic radiation, which produces sharp peaks at certain energy locations—for instance, see Figure 1 in Section 3 for a typical x-ray spectra. The challenge is to recover both structures simultaneously, so that the estimated spectrum accurately represents the spectral response of the x-ray imaging system.

In this work, we exploit prior knowledge of the x-ray spectrum to design a suitable regularizer when we formulate an optimization problem; in this way, we can allows for recovering both structures and at the same time overcome the ill-conditioning of the problem.

## 2.2   EM method

Expectation-maximization (EM) is a widely used optimization method, which has frequently been applied to the problem of x-ray spectrum estimation (e.g. Sidky et al. [12], Duan et al. [18]). Broadly speaking, EM is a general framework for solving a maximum likelihood estimation problem when the obtained data is incomplete. In the setting of spectrum determination, the incompleteness of the data arises from the fact that the detected photon count along ray $\ell$ is observed through a sum of transmitted photon counts across frequencies $i$, namely, $c_\ell \approx \sum_i X_{\ell i} s_i$ for the system matrix $\{X_{\ell i}\}$. Under the Poisson noise, EM then finds the maximum likelihood estimate $\{\widehat{s}_i\}$,

$$\widehat{s} = \arg \min_s \left[ \sum_\ell \left\{ \sum_i X_{\ell i} s_i - c_\ell \log \left( \sum_i X_{\ell i} s_i \right) \right\} \right],$$

by applying the following iterations

$$s_i^{(t+1)} = \frac{s_i^{(t)}}{\sum_\ell X_{\ell i}} \sum_\ell \frac{X_{\ell i} c_\ell}{\sum_{i'} X_{\ell i'} s_{i'}^{(t)}} \text{ for all } i. \tag{2}$$

Here the update equation is derived by minimizing the EM objective function $Q(s; s^{(t)})$, given by

$$Q(s; s^{(t)}) = \sum_{\ell i} \left\{ X_{\ell i} s_i - c_\ell \frac{X_{\ell i} s_i^{(t)}}{\sum_{i'} X_{\ell i'} s_{i'}^{(t)}} \log(X_{\ell i} s_i) \right\}. \tag{3}$$

Note that the multiplicative form of the update equation (2) automatically guarantees non-negativity of the solution as long as the initial value is set to non-negative.

For an underdetermined and noisy linear system, the maximum likelihood solution is known to overfit to the data and yield undesirable structure of the x-ray spectrum. Hence, it is desirable to minimize the data discrepancy function (the transmission Poisson likelihood function in this case) but possibly subject to constraints and/or regularization on the spectrum. Meanwhile, it is known that the iterations (2) are guaranteed to converge to the solution $\{\widehat{s}_i\}$ for any initial point. Therefore, if run to convergence, the EM iterations should reach the solution $\{\widehat{s}_i\}$ and thus fail to deliver accurate estimation of the x-ray spectrum. In particular, if we try to incorporate prior information about the x-ray spectrum into the initialization of EM, we would expect it to still end up at the same solution, $\{\widehat{s}_i\}$. To avoid this issue, early stopping of EM is often employed (e.g. Sidky et al. [12]) to regularize the algorithm path and avoid overfitting to the data. Note that, due to the global convergence property of EM, the idea of incorporating prior information via initialization makes sense only in the context of early stopping.

While EM enjoys many empirical advantages for spectrum reconstruction, our motivation to derive a spectrum reconstruction method from an optimization framework is to enhance interpretability and flexibility of the reconstruction procedure; for EM, it is not clear what kind of regularization early stopping is performing for the algorithm, and other desirable constraints on the spectrum, such as a normalization constraint, cannot be easily incorporated. On the other hand, our framework is capable of including multiple constraints on the spectrum, and moreover, we do not require any form of early stopping but rather fully minimize target optimization problem for accurate reconstruction of x-ray spectrum. Our approach might also allow us to build towards simultaneous spectrum estimation and basis material maps reconstruction in spectral CT.

### 2.3   Spectrum estimation via KL-divergence constraint

Now we turn to the development of our method to estimate the x-ray spectrum from transmission measurements through an optimization problem.

We assume that an initial spectrum, namely a prior estimate of the x-ray spectrum, is available such that the initial value exactly captures the characteristic peaks of the target spectrum (without such information, we cannot hope to recover details of the spectral curves such as the characteristic peaks.) Denoting the initial value by $\{s_i^{\text{ini}}\}$, we measure the distance of the x-ray spectrum $\{s_i\}$ to the initial value via Kullbeck-Leibler (KL) divergence, i.e. $d_{\text{KL}}\left(s, s^{\text{ini}}\right)$, where for positive vectors $x \geq 0, y > 0$, the KL divergence is defined by

$$d_{\text{KL}}\left(x, y\right) = \sum_i \left\{x_i \log(x_i/y_i) + y_i - x_i\right\}.   \tag{4}$$

(Note that the definition (4) reduces to the usual definition of KL-divergence over probability vectors, when the vectors $x, y$ satisfy $\sum_i x_i = \sum_i y_i = 1$.) The KL-divergence is convex in $(x, y)$ and satisfies $d_{\text{KL}}\left(x, y\right) \geq 0$ for $x \geq 0, y > 0$, and $d_{\text{KL}}\left(x, y\right) = 0$ if and only if $x = y$. In order to stabilize inversion of the data model, a KL-divergence constraint, i.e. a bound on $d_{\text{KL}}\left(s, s^{\text{ini}}\right)$, is placed on the estimated x-ray spectrum $\{s_i\}$ to control the deviation from the initial value.

Specifically, the x-ray spectrum is reconstructed through the following constrained minimization problem:

$$\begin{aligned} \underset{s}{\text{minimize}} \quad & d_{\text{KL}}\left(c, Xs\right) \\ \text{subject to} \quad & d_{\text{KL}}\left(s, s^{\text{ini}}\right) \leq c, \\ & \sum_i s_i = 1, s_i \geq 0 \text{ for all } i, \end{aligned}   \tag{5}$$

for a constraint parameter $c \geq 0$, where the KL-divergence is employed for both the data discrepancy function and the constraint function. Note that the data discrepancy function here, namely, KL-divergence between measured data and calculated photon counts, is equivalent to the transmission Poisson likelihood (TPL) function up to constant terms (Barrett and Myers [19]) hence the solution of the problem (5) is equivalent to a constrained maximum likelihood estimate of the counts data under a Poisson noise assumption. The TPL function can be useful even when the measured counts data is inconsistent with the Poisson assumption, since it assigns more weight to higher count measurements (Barber et al. [17]). The $\ell_1$-norm constraint, $\sum_i s_i = 1$, ensures normalization of the resulting solution, which endows physical meaning to the reconstructed x-ray spectrum.

Although the description of the data model (1) is idealized, the proposed optimization-based approach is flexible and can include other physical effects, such as x-ray scatter as well as other non-ideal detector effects, in the estimation process by adding constraints or modifying the objective function. The use of KL-divergence as a constraint function can be valid for any given optimization formulations. In terms of computation, the problem (5) is a convex program, so any convex solver can be applied to solve the problem efficiently. For instance, we have implemented the method using the "cvx" package in Matlab with solver MOSEK which solves the problem (5) in less than a second. Alternatively, we can apply the exponentiated-gradient (EG) algorithm, which is a simple first-order algorithm that iteratively performs a descent step followed by projection onto the feasible region of x-ray spectra. See Section 2.5 for a detailed discussion of the EG algorithm and convergence guarantees for obtaining optimal solution of the problem (5).

Care must be taken in specifying the initial value $\{s^{\text{ini}}\}$ as it has a great influence on the final estimation of the x-ray spectrum. If the employed initial value reflects realistic structure of a spectral curve, the resulting solution can provide accurate estimation of the target spectrum and therefore accurately reproduce transmission measurements. The robustness of the method with respect to the initial spectrum is also investigated in the simulation study (see Section 3.1).

### 2.4   Connection to maximum entropy method

The proposed method based on KL-divergence is closely related to the well known *principle of maximum entropy* in the existing literature. This principle state that, of all possible solutions that are consistent with the data, we choose the one with the largest entropy $-\sum_i s_i \log s_i$, or with the least divergence (or relative entropy) $\sum_i s_i \log(s_i/s_i^{\text{ini}})$ if the prior information $\{s_i^{\text{ini}}\}$ is known. The maximum entropy principle has been widely studied in the following decades, with applications to a broad range of problems including image reconstruction from incomplete and noisy data (Gull and Daniell [20]). We refer the reader to Shore and Johnson [21] for justification of the principle.

In the context of spectrum estimation, applying the maximum entropy principle with prior information $\{s_i^{\mathrm{ini}}\}$ leads to the following constrained optimization problem:

$$
\begin{aligned}
\underset{s}{\text{minimize}} \quad & \mathrm{d_{KL}}\left(s, s^{\mathrm{ini}}\right) \\
\text{subject to} \quad & \mathrm{d_{KL}}\left(c, Xs\right) \leq C, \\
& \sum_i s_i = 1, s_i \geq 0 \text{ for all } i,
\end{aligned}
\tag{6}
$$

where we again employ the TPL discrepancy function as a measure of the fit to the data, and $C > 0$ is a parameter that limits the amount of this discrepancy.

Now, since the problem is convex in the variable $\{s_i\}$, we can find a one-to-one correspondence between the parameters $c$ in (5) and $C$ in (6) such that the solutions from both optimization problems exactly match; this, in turn, implies that the problem (5) is equivalent to the problem (6), and particularly shows the equivalence between the proposed approach and the maximum entropy principle. This provides a justification of the use of KL-divergence as a constraint function for spectrum estimation. On the other hand, note that the convexity of the TPL discrepancy function is essential here. While the KL-divergence constraint can be applied to the data models including other physical factors, the resulting data discrepancy function can generally be nonconvex in which case the equivalence property is no longer guaranteed to hold. Even in such case, however, we believe that a similar kind of interpretation can be useful in gaining insight into the constrained approach with KL-divergence.

## 2.5  Exponentiated-gradient algorithm

While the problem (5) can generally be solved by any convex solver, in some applications, it is useful to have an iterative algorithm that solves the problem more explicitly. In this work, we solve this optimization problem using the exponentiated-gradient (EG) algorithm (Kivinen and Warmuth [22]), that is designed to solve general convex objectives over the simplex $\{s : \sum_i s_i = 1, s_i \geq 0 \text{ for all } i\}$. Exponentiated-gradient algorithm can also be viewed as a special case of mirror descent with the mirror map given as the negative entropy function (Bubeck [23]).

First, we write the constrained problem (5) in the equivalent Lagrangian form:

$$
\begin{aligned}
\underset{s}{\text{minimize}} \quad & \mathrm{d_{KL}}\left(c, Xs\right) + \lambda \cdot \mathrm{d_{KL}}\left(s, s^{\mathrm{ini}}\right) \\
\text{subject to} \quad & \sum_i s_i = 1, s_i \geq 0 \text{ for all } i,
\end{aligned}
\tag{7}
$$

where $\lambda$ is a regularization parameter that controls the amount of regularizing effect, and the constraints represent the feasible region of x-ray spectra. Again, there is a one-to-one correspondence between $c$ in (5) and $\lambda$ in (7), due to the convexity of the problem.

The EG algorithm applied to the above problem yields the following iterations: initialize $s^{(0)} = s^{\mathrm{ini}}$, fix the step size $\eta > 0$, then for steps $t = 0, 1, 2, \ldots,$

$$
\begin{cases}
\text{Set } g^{(t)} = \nabla_s \mathrm{d_{KL}}\left(c, Xs^{(t)}\right) + \lambda \nabla_s \mathrm{d_{KL}}\left(s^{(t)}, s^{\mathrm{ini}}\right); \\
\text{Set } s_i^{(t+1)} \leftarrow s_i^{(t)} \exp\left(-\eta \cdot g_i^{(t)}\right) \text{ for all } i; \\
\text{Set } s^{(t+1)} \leftarrow s^{(t+1)} / \|s^{(t+1)}\|_1.
\end{cases}
\tag{8}
$$

Now examining the steps given in (8), we see that the update equation of $s_i^{(t+1)}$ is multiplicative as analogous to the EM iterations (2). Particularly, this guarantees automatic inclusion of non-negativity constraints in the estimated spectrum, as long as the initial spectrum is non-negative. On the other hand, a distinct feature of the EG algorithm is that at every iteration the normalization constraint is enforced by the projection step $s^{(t)} / \|s^{(1)}\|_1$ (more precisely, the projection is performed with respect to the KL-divergence), whereas the EM method can give no such guarantees on the final solution. The projection step can be optional, and is not needed if the normalization constraint is not included in (7). To compare the EG and EM algorithms, while the EM algorithm seeks to minimize (3) at each iteration to reach the maximum likelihood solution (if EM is run to convergence), EG instead seeks to take each step that monotonically decreases (3) with additional KL-divergence regularization term. Both algorithms will produce a sequence of estimates that will decrease the (regularized) data discrepancy at each iteration.

The convergence of the EG algorithm has been well established in the literature—for instance, it is shown that the objective gap between the point at iteration $t$ and the optimal solution decays with the rate $\mathcal{O}\left(\frac{1}{t}\right)$ (Bubeck [23]). In Appendix A, we provide a simple way to test for convergence by checking whether the KKT conditions is satisfied within a predefined threshold $\epsilon > 0$ (Arora et al. [24]). Implementing the algorithm is quite straightforward, but one needs to specify the step size $\eta > 0$ at every iteration. In general, the step size can be chosen to be fixed or with a line search method. For the numerical experiments studied in this work, we perform the EG algorithm with a fixed step size, for which the algorithm is observed to converge rapidly to the optimal solution. Details for our implementations of the algorithm can be found in Section 3.

# 3 Results

## 3.1 Simulation study

Now we perform a numerical experiment on the simulated transmission measurements to examine the empirical performance of the proposed method, as well as compare to the EM method.

A step wedge phantom is modeled and simulated, consisting of Aluminum and Polymethyl Methacrylate (PMMA). The thicknesses of Aluminum and PMMA are each selected in the range of $\{0, 0.635, 1.270, 1.905, 2.540\}$ and $\{0, 2.540, 5.080, 7.620, 10.160\}$ respectively, giving a total of 25 combinations across the step wedge. The linear attenuation coefficients are obtained using the NIST table (Berger et al. [25]). Three kinds of polychromatic spectra, sampled at 1 keV intervals between 10 keV and 100 keV, are employed to either generate transmission measurements, or to serve as an initial value for the effective spectrum estimation; those spectra are determined from the experimental data described in Section 3.2, and represent a typical spectral response of the photon-counting CT system for energy windows with thresholds at 25 keV, 40 keV, and 60 keV. Using the experimentally determined spectra allows us to model the rational shape of the x-ray spectrum.

Given the true spectrum, the expected total transmitted photon counts $\{\hat{c}_\ell\}$ are computed according to the data model (1) with expected incident photon counts $N_\ell = 10^5$ for each ray $\ell$. The noisy measurements $\{c_\ell\}$ are then generated with an independent Poisson model from which the true x-ray spectrum is reconstructed. Additionally, we generate another set of noisy transmission measurements through 20 different thicknesses of water which are varied from 0 to 20 centimeters at equal intervals, and where the NIST values are used to obtain the energy dependent attenuation coefficients. These measurements are not included in the reconstruction of the x-ray spectrum, but will serve as a "validation" set to assess the reproducibility of the spectrum estimation methods.

The x-ray spectrum is reconstructed by solving the optimization problem (7) with an implementation of the EG algorithm, as described in Section 2.5. Recall that $\lambda$ is the user-defined parameter to control the trade-off between the data fidelity of the model and the regularization on the KL-divergence of the solution. We vary $\lambda$ over $\lambda \in \{20, 30, \ldots, 1000\}$, and select the value that minimizes the root mean square error (RMSE)

$$\text{RMSE}(\lambda) = \sqrt{\frac{\sum_i (s_i(\lambda) - s_i^{\text{true}})^2}{\sum_i (s_i^{\text{true}})^2}}, \tag{9}$$

where $\{s_i(\lambda)\}$ is the estimated spectrum given this choice of $\lambda$, and $\{s_i^{\text{true}}\}$ is the true spectrum. The spectrum achieving the minimum RMSE will be close to the true spectrum in shape, and thus can reliably reproduce transmission curves for any configurations of materials. For step size, we fix $\eta = 1.3 \cdot 10^{-5}$ throughout the simulation. We run the EG algorithm (8) until convergence, where we check the convergence of the algorithm as given in Appendix A. For the present work, we set the threshold $\epsilon = 10^{-8}$.

Figure 1(a,b) show the spectral curves reconstructed from transmission measurements by employing the ground truth and initial spectrum shown in the figures, respectively. For each given ground truth and initial spectrum, we simulate 20 independent sets of transmission measurements and obtain the best spectrum solutions by running the EG algorithm. Hence, each plot of Figure 1(a,b) shows reconstructed x-ray spectrum for 20 different sets of measurements. Due to the noise, there exist some variation between the spectral curves. As seen in the figures, however, the spectra generated

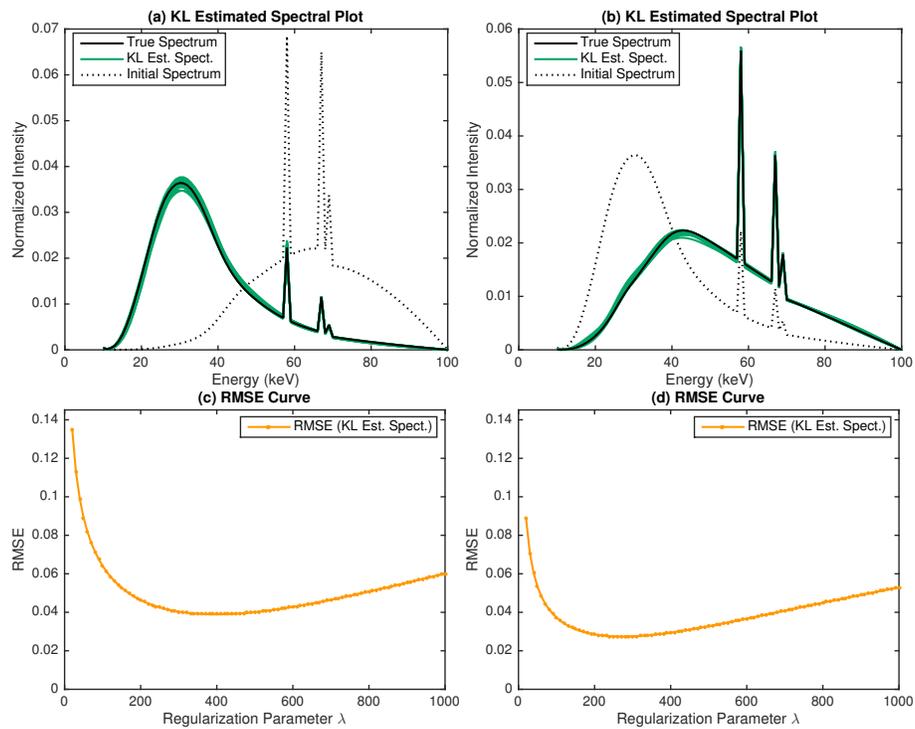

Figure 1: Spectrum estimation from simulated transmission measurements by use of KL. Different types of true and initial x-ray spectrum are employed shown in black solid line and black dotted lines respectively. In each setting, spectrum reconstruction is performed for 20 independent sets of transmission measurements. Panels (a),(b): Spectral curves for 20 different trials. The band formed by the curves shows variation between the reconstructed x-ray spectra. Panels (c),(d): The RMSE curves computed by (9) for different regularization parameters. Each point represents an average over 20 trials.

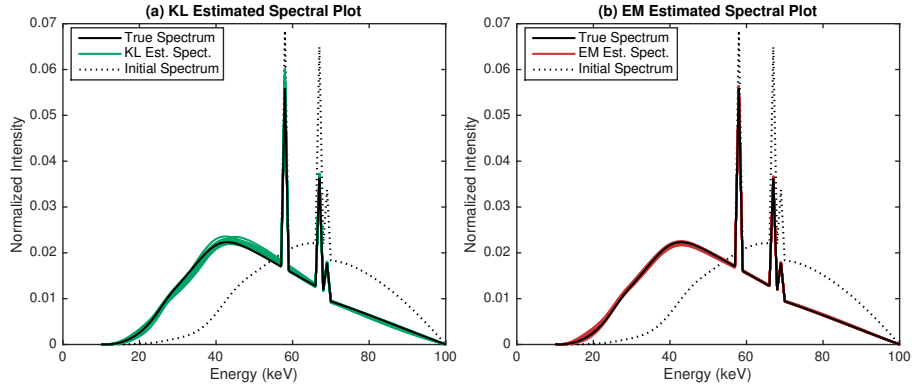

Figure 2: Comparison of spectrum estimation from simulated transmission measurements by use of KL and EM. Results for spectral curves, fitted by KL and EM respectively, are displayed for 20 different trials.

by the method are concentrated near the respective true spectra and furthermore every single spectrum resembles the shape of its target with a high precision. More importantly, the results further show the robustness to the shapes of the chosen ground truth and initial spectra; the method continues to perform well, as long as the initial spectrum shares the same locations of the characteristic peaks as the true spectrum even though the relative intensities can be substantially different. This property can be particularly favorable for spectral calibration of a photon-counting detector, since spectral information from one energy window can be useful for estimating the spectral response of other windows.

The lower row of Figure 1 displays the RMSE plots, averaged over 20 trials, with respect to regularization parameter $\lambda$. For the two plots, the method yields larger error at first, but drops rapidly thereafter and achieves a minimum at $\lambda$ in the range of 200–400. The error remains relatively lower in a broad range of $\lambda$'s around the minimum, which illustrates that the method is numerically stable relative to the choice of $\lambda$. At larger values of $\lambda$, bias is induced in the solution and the error from the true spectrum begins to grow again. In comparison to the other case, the RMSE curve is placed higher in Figure 1(c), which results from the fact that the employed initial spectrum is farther from the truth than the other case.

Figure 2(a) and (b) show comparison of the spectra fitted by the KL-divergence based method and the EM method from simulated transmission measurements. For EM, the number of iterations is varied from 10 to $10^4$ and the optimal number is chosen based on the RMSE rule described in (9). While it is seen that EM tends to estimate the true spectrum more faithfully (the averaged RMSE values by the best case KL and EM solutions are 0.0350 and 0.0184 respectively), the spectrum representations by both methods generally exhibit comparable performance in recovering physical shape of the true spectrum. Moreover, the utility of the KL-divergence approach lies in the mathematical formulation of spectrum estimation as an optimization problem.

Next, we evaluate the prediction of the transmission curves using the spectrum estimates based on water transmission measurements at 20 thicknesses. We use the $\ell_2$-distance for log counts

$$\sum_{\ell'} (\log(c_{\ell'}) - \log(\hat{c}_{\ell'}))^2 \tag{10}$$

to measure the prediction performance. Figure 3(a) shows the prediction error of the KL-divergence approach plotted against the varying regularization parameter, as well as the prediction by the best case EM solution (which, recall, minimizes the RMSE criterion in (9)) and the true spectrum for reference (note that even the true spectrum cannot perfectly reproduce the transmission data due to the noise). For small values of $\lambda$, the KL-divergence approach performs nearly as well as the best case EM solution and slightly less than the true spectrum, demonstrating its capability to represent the measurement process; for higher values of $\lambda$, however, the performance rapidly degrades which results from the underfitting of the model. Figure 3(b) displays the actual transmission curves predicted by both methods, as well as the simulated water transmission data and the transmission curve predicted by the initial spectrum. Without loss of generality, here we only give a representative result from different trials. Again it is clearly seen that both predicted transmission curves are accurate enough to predict the water transmission data and show the significant improvement over the transmission curve predicted by the initial spectrum.

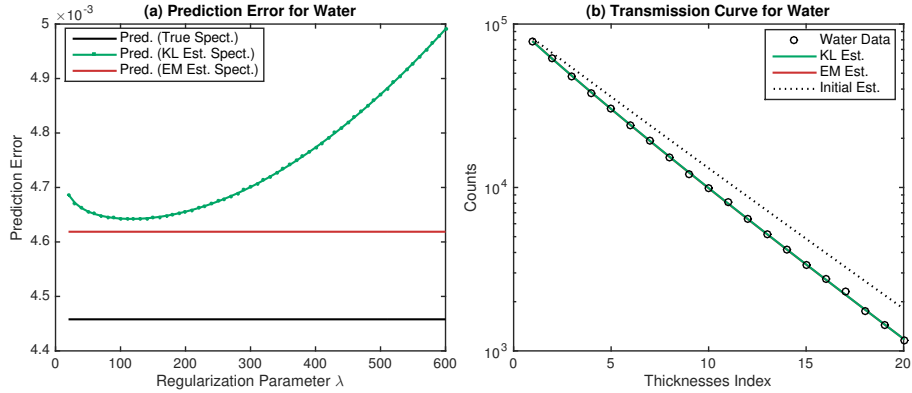

Figure 3: Panel (a): Prediction error in the transmission curves derived from the x-ray spectra using KL and EM. For the EM error, the best case solution is used to produce the transmission curve, which is irrelevant with respect to the regularization parameters. Each point represents an average over 20 trials. Panel (b): Plot of the predicted transmission curves for the reference material water. The x-axis indicates the thicknesses index $\ell'$ for water, and the y-axis is plotted on a logarithmic scale. The results for KL and EM are nearly identical and cannot be distinguished in the plot.

## 3.2 Experimental study

The proposed KL-divergence approach is evaluated on the experimental data which is performed on a bench-top x-ray system consisting of a microfocus x-ray tube and a photon-counting Cadmium-Zinc-Telluride (CZT) detector comprised of 128 detector pixels, of which 96 are usable. A step wedge phantom made of Aluminum and PMMA, shown in Figure 4(a), are measured at the same dimensions with the simulated step wedge as described in Section 3.1. We refer to our coauthors' paper (Schmidt et al. [6]) for more details of the experimental setup.

The initial spectrum is generated with the SPEC78 software from the IPEM78 report (Cranley et al. [26]), which contains the expected spectrum exiting the tube for a 100 kV beam with 1-mm of aluminum filtration. Based on the measurement sets, reconstruction is performed with the KL-divergence regularized problem (7) to estimate effective spectral response of the photon-counting detectors for each energy window and detector pixel. Determining a good regularization parameter is critical in obtaining an accurate x-ray spectrum. The RMSE rule (9) cannot be applied here, since the true spectrum is unknown in the experimental setting. A validation method is another attractive option to choose a good value of $\lambda$, for which we randomly partition the transmission measurements into the training data and test data and select $\lambda$ that best predicts the test data using the x-ray spectrum reconstructed from the training data. While the validation method is observed to perform well in the simulation setting, we find that when applied to the experimental setting, the estimated spectra tend to highly overfit the experimental data and show unphysical fluctuations in the resulting curves. This is attributed to the systematic dependencies present in the measured photon counts, which can arise from various non-ideal physical effects of photon-counting detectors that have not been included in the data model (1).

For the current experiment, we instead rely on *ad hoc* procedure for selecting the optimal value of $\lambda$. The selection rule is based on the observation that the bremsstrahlung spectrum typically reveals unimodal structure in the corresponding energy region. The initial spectrum, shown in Figure 4(b), exhibits characteristic peaks at $58, 67, 69$ keV, but in other regions, the curve is smooth and nearly unimodal—it has a local minimum at $s_{11}^{\text{ini}}$ (not visible in the figure), and a local maximum at $s_{33}^{\text{ini}}$. We expect to see this type of simple structure in the true spectrum as well. We therefore choose regularization parameters $\lambda$ that yield the spectrum whose bremsstrahlung part reflects the same unimodal structure as the initial spectrum. More specifically, consider the spectrum $\{s_i\}$ constrained to the bremsstrahlung part of the frequency curve, by removing the characteristic peaks at $58, 67, 69$ keV:

$$s_{\text{brem}} = (s_{10}, \ldots, s_{57}, s_{59}, \ldots, s_{66}, s_{68}, s_{70}, \ldots, s_{100}),$$

i.e. the energy spectrum of photons is decomposed into $s_{\text{brem}}$ and $s_{\text{char}} = (s_{58}, s_{67}, s_{69})$. We choose the regularization parameter by taking the smallest value of $\lambda$ such that the estimated spectrum, $s(\lambda)$, exhibits at most one local minimum and one local maximum, when the characteristic peaks are removed—that is, at most one local minimum and one local maximum in the vector $(s(\lambda))_{\text{brem}}$, the bremsstrahlung part of the estimated spectrum. We expect that values of $\lambda$ which are too small, leading to insufficient regularization, would yield an estimated spectrum $s(\lambda)$ that overfits to the data, which would typically exhibit many local minima and maxima; therefore, our procedure ensures that we choose a value of $\lambda$ that is not too small, to avoid overfitting.

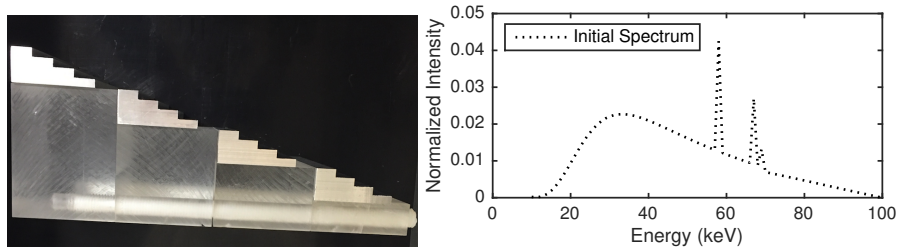

Figure 4: Left: (a) Step wedge phantom used for spectral calibration in x-ray imaging. Right: (b) The initial x-ray spectrum.

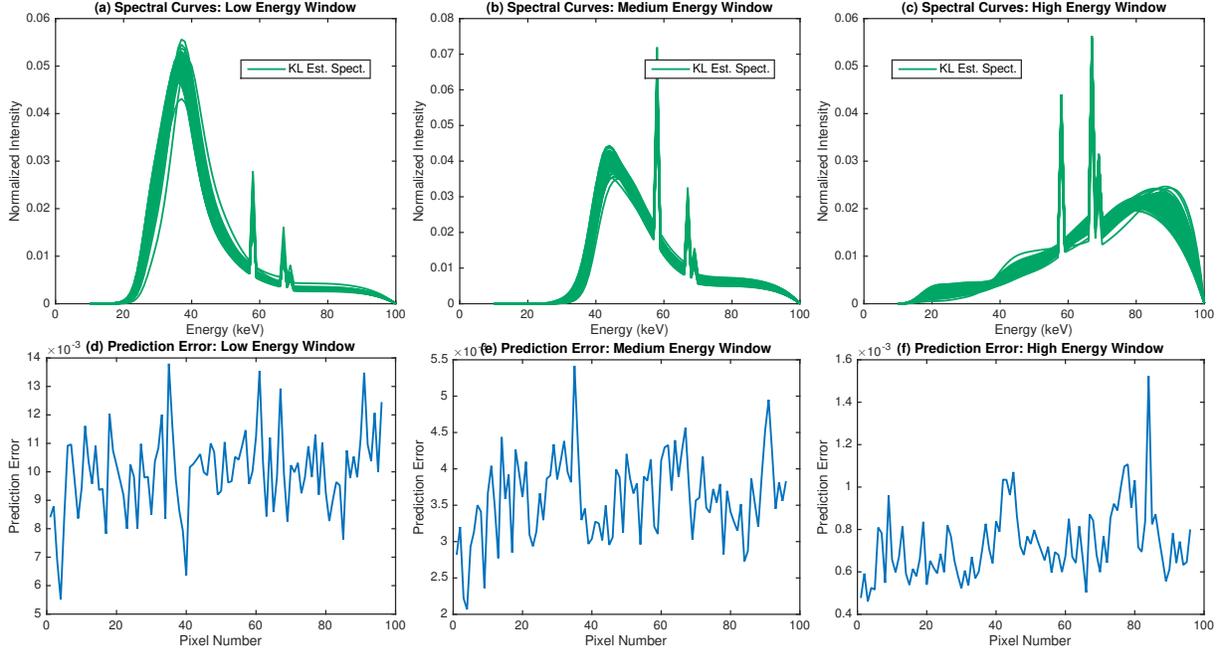

Figure 5: Spectrum estimation from the measured transmission data by use of KL. Each column represents the results for different spectral windows. Panels (a),(b),(c): Spectral curves for each energy window. The plots show the solution curves for 96 different detector pixels. Panels (d),(e),(f): Prediction error in the transmission curves derived from the x-ray spectra shown above across detector pixels.

Results for experimental data are shown in Figure 5. Each panel in the upper row shows the reconstructed x-ray spectra for three different energy windows, as well as the initial spectrum depicted as the dotted line. Within each panel, the curves are obtained by running the EG algorithm from 96 different detector pixels, where step size is set to $\eta = 1.3 \cdot 10^{-5}$. While there is substantial variation in the reconstructed x-ray spectra across the detector pixels, the selection method based on the unimodality consistently yields spectra that resemble realistic shapes of the bremsstrahlung and characteristic lines. Compared to the results for high energy window, the spectra estimated for low and medium energy windows appear to follow the realistic shape more faithfully. The spectral curves displayed in high energy window seem to be less stable and exhibit more fluctuations in the bremsstrahlung region. Similar results are also observed by comparing the prediction errors for different energy windows shown in the lower row of Figure 5, where it is suspected that the method tends to overfit to the data for high energy window in comparison to the other windows. Of course we can increase the penalization parameter $\lambda$ to avoid this problem of overfitting, but the resulting spectra will now be strongly biased towards the initial spectrum. In principle, the problem of calibrating spectral response for high energy window is more difficult than the other cases, because the consecutive photons with low energies can be wrongly counted as the single photon with high energy, leading to a degradation of the spectral measurements in the high energy window.

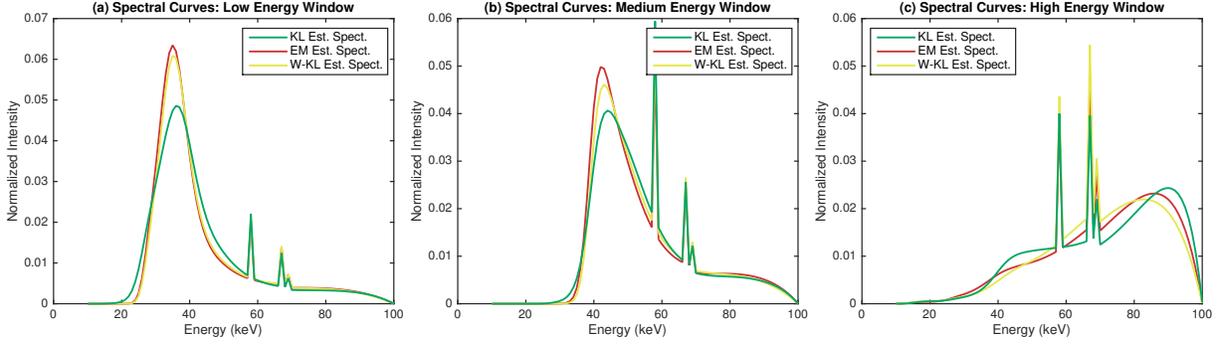

Figure 6: Comparison of spectrum estimation from the measured transmission data by use of KL, EM, and weighted KL. Results are shown for one particular detector pixel (pixel number = 34). Each panel shows the reconstructed spectra for different spectral windows, along with the initial spectrum.

To improve the stability of the estimated spectra for high energy window, we implement a simple variant of the proposed method that imposes KL-divergence regularization on the spectrum with different weights on each component of the spectral density $s_i$. In particular, we solve the following regularized optimization problem:

$$\begin{aligned} \underset{s}{\text{minimize}} \quad & \text{d}_{\text{KL}}\left(c, Xs\right) + \lambda \cdot \text{d}_{\text{KL}}^{\text{w}}\left(s, s^{\text{ini}}\right) \\ \text{subject to} \quad & \sum_i s_i = 1, s_i \geq 0 \text{ for all } i, \end{aligned} \tag{11}$$

where $\text{d}_{\text{KL}}^{\text{w}}\left(x, y\right)$ is a weighted KL-divergence given by

$$\text{d}_{\text{KL}}^{\text{w}}\left(x, y\right) = \sum_i w_i \left\{ x_i \log\left(x_i/y_i\right) + y_i - x_i \right\},$$

for a weight vector $\{w_i\}$. In the current setting, each column of the system matrix $\{X_{\ell i}\}$ that contributes to the measured photon counts has different scalings, so we choose to use the weights such that $w_i \propto \sum_\ell X_{\ell i}$ for each $i$. This helps to treat different spectral densities $s_i$ on a more equal basis. The EG algorithm can also be applied to solve the problem (11).

Figure 6 shows spectral curves reconstructed by the three methods, the original KL-divergence based method, its weighted version given in (11), and the EM method, from the measured counts data for different spectral windows. Here we fix the detector pixel (pixel number = 34) such that the spectrum returned by the KL-divergence approach exhibits some fluctuation in the high energy window. We can see that employing the weighted KL-divergence removes such unphysical shape in the resulting curve and makes the spectrum more smooth in the bremsstrahlung energy region. Moreover, it is interesting to see that in all energy windows, the weighted KL-divergence and the EM method yield x-ray spectra that are close in shape, but have some deviations from the x-ray spectra generated by the KL-divergence based method. We observe this phenomenon not only for the measured data at this particular detector pixel, but across all detector pixels. This is in sharp contrast to the results shown in simulation study where both the KL-divergence approach and the EM method yield x-ray spectra that closely resemble the ground truth. Under the presence of inconsistency between the data model (1) and the physical transmission model, the KL-divergence based method can perform quite differently in comparison to EM and the weighted KL-divergence approach.

In Figure 7, the prediction performance is evaluated using the fitted x-ray spectra shown in Figure 6, where the error is computed according to the squared log count distance (10). We can see that all three methods significantly improve the prediction of the transmission curves compared to the initial spectrum. The residuals between the measured and predicted transmission curves are shown in the lower row of Figure 7. While the residuals generally behave similarly between the three methods, in the case of low and medium energy windows, the EM method generates larger residual errors for small thicknesses indexes; this is attributed to the fact that spectrum normalization constraint is not imposed in the EM solutions, which leads to errors in the transmission curves when there is no object in the scan system (thickness index 1 in the figure corresponds to the absence of an object). For high energy window, the three methods appear to perform similarly in terms of predicting the transmission data, though the curves between the KL-divergence approach and EM are more similar than the weighted KL-divergence. It is also worth noting that the plots displayed in Figure 7 clearly show visible trends in the residual errors, indicating the presence of systematic errors due to the unmodeled physics in the measurement process. In particular, this suggests the need for employing more realistic modeling of physical factors in order to account for the limitation of the method and enable more accurate and effective spectrum estimation.

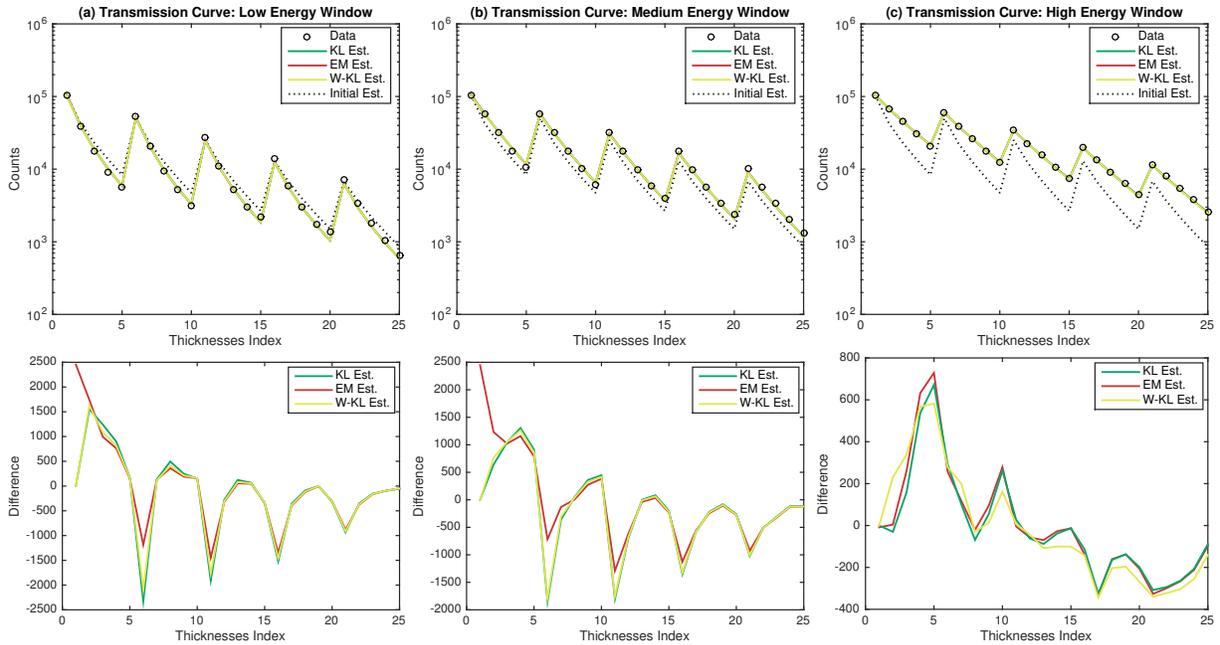

Figure 7: Comparison of spectrum estimation from the measured transmission data by use of KL, EM, and weighted KL. Results are shown for one particular detector pixel (pixel number = 34). The x-axis indicates the thicknesses index $\ell$ for the step wedge. The upper row of each panel shows prediction error in the transmission curves derived from the x-ray spectra shown in Figure 6. The lower row of each panel shows the residuals between the measured transmission curve and the predicted transmission curves.

# 4    Discussion and Conclusions

In this paper, we have developed a constrained optimization problem for reconstructing x-ray spectrum from transmission measurements through known thicknesses of known materials. The proposed method places a KL-divergence constraint on the spectrum variable which improves numerical stability of the inversion process and allows to incorporate prior knowledge on the spectrum. The formulated optimization problem is a convex program over the simplex, which we propose to solve based on the exponentiated-gradient algorithm. Both numerical simulations and experimental results show that the method can yield realistic x-ray spectra that can accurately reproduce the spectral response of the CT system. Furthermore, a simple variation of the method that employs weighted KL-divergence appears to perform well in the experimental data when the measurements are highly noisy and inconsistent with the data model.

In realistic applications, the measured data is affected by many physical factors such as x-ray scatter and various physical processes involved in a photon-counting detector. The simple data model assumed in this work is not sufficient to describe the realistic measurement process of the imaging system, and some corrections and extensions are required to further improve the quantitative accuracy of the reconstructed spectrum. For the x-ray scatter effects, one can perform scatter correction for spectrum estimation on the EM method (Lee and Chen [27]). An analogous approach can be considered to combine with the method presented in this work.

While the numerical experiments demonstrate a comparable performance with the proposed method and EM, we emphasize the other benefits of our algorithm relative to EM. First, the proposed approach using a KL-divergence constraint is a general optimization framework for spectrum estimation that supports different data discrepancy functions and can easily incorporate other desirable constraints on the x-ray spectrum. The data model described in (1) can be generalized to take the non-ideal detector physics into account in the acquisition of the photon counts. By formulating the inversion of the data model into the optimization problem, the parameters specifying the detector effects can be estimated within our framework of spectrum estimation. Besides the flexibility of the method, our formulation also provides the advantage in terms of interpreting spectrum determination from transmission measurements. In particular, under the assumption of the simple data model (1) and using the TPL discrepancy function, the equivalence to the maximum entropy method is established which enriches our understanding of the spectrum estimation problem in connect to the long history of the maximum entropy literature. The flexibility and interpretability of the proposed method makes it promising for spectrum determination in practical spectral calibration applications.

Finally, incorporating the proposed optimization calibration procedure in the framework of simultaneous spectral calibration and spectral CT image reconstruction can be an interesting future research direction. In previous work on spectral CT image reconstruction, we have incorporated unknown spectral response scaling factors in the spectral CT data model and we performed simultaneous image reconstruction and estimation of these scaling factors (Schmidt et al. [6], Barber et al. [17]). This approach allowed for the reduction of ring artifacts in the reconstructed images. Investigating the possibility of combining with the KL-divergence for imposing constraint on the spectral components can potentially be useful for auto-calibration of the spectral response of the imaging system during the spectral CT image reconstruction.

# Acknowledgements

R.F.B. is supported by an Alfred P. Sloan Fellowship and by NSF award DMS-1654076. T.G.S. is supported by NIH R21EB015094. This work is also supported in part by NIH Grant Nos. R01-EB018102, and R01-CA182264. The contents of this article are solely the responsibility of the authors and do not necessarily represent the official views of the National Institutes of Health.

# A    Convergence test

The convergence test of the exponentiated-gradient algorithm has appeared in the literature (Arora et al. [24]), which we provide here for the sake of completeness. The Lagrangian function associated with the problem (7) is given by

$$\mathcal{L}(s, \nu, \gamma) = \mathrm{d_{KL}}\left(c, Xs\right) + \lambda \cdot \mathrm{d_{KL}}\left(s, s^{\mathrm{ini}}\right) - \sum_i \nu_i s_i + \gamma \cdot (\sum_i s_i - 1).$$

By the KKT condition, the optimal solution satisfies the following conditions:

(a) $\sum_i s_i = 1$ and $s_i \geq 0\ \forall i$.

(b) $\nu_i \geq 0\ \forall i$.

(c) $\nu_i s_i = 0\ \forall i$.

(d) $\nabla_s \mathcal{L}(s, \nu, \gamma) = 0$.

Set $\gamma = -\min(\nabla_s \mathrm{d_{KL}}\left(c, Xs\right) + \lambda \nabla_s \mathrm{d_{KL}}\left(s, s^{\mathrm{ini}}\right))$ and $\nu = \nabla_s \mathrm{d_{KL}}\left(c, Xs\right) + \lambda \nabla_s \mathrm{d_{KL}}\left(s, s^{\mathrm{ini}}\right) + \gamma \cdot \mathbf{1}$, where $\min$ is taken componentwise. Then it can be checked that the conditions (b),(d) are satisfied. Also the condition (a) is trivial since the optimal solution is always feasible from the update equation (8). It remains to check the complementary slackness condition (c). By the conditions (a),(b), we know that $\nu_i \cdot s_i$ is non-negative, so the condition (c) is implied if $\sum_i \nu_i s_i = 0$. Therefore, we can test convergence of the algorithm by checking $\sum_i \nu_i s_i < \epsilon$ for a predefined threshold $\epsilon > 0$.

# B    Additional simulation results

Here, we provide simulation results that investigate the effect of early stopping on the x-ray spectrum reconstruction using the EM algorithm (2). Figure 8 shows the spectral curves estimated by EM for different numbers of iterations. The transmission data are simulated as described in the simulation study (Section 3.1). As seen in the figure, by stopping after 300 iterations, the EM method recovers the ground truth spectrum remarkably well and both spectra are indistinguishable in the plot; however, for low iteration number (e.g. 10 iterations), the resulting spectrum is still biased towards the initial value, and for high iteration number (e.g. 5,000 iterations), the EM method appears to overfit to the transmission data and therefore cannot generalize to transmission measurements beyond the given data set. While determining a good iteration number is crucial to implement the EM method, Sidky et al. [12] demonstrates the robustness of the EM method, i.e. the estimated x-ray spectrum is not strongly sensitive to the choice of number of iterations. In practice, the number of iterations can be determined by preliminary simulation studies.

Next, we investigate how the choice of the initial spectrum affects on the performance of the proposed KL-divergence based method. We fix the ground truth spectrum as depicted in Figure 9(a) with black solid line, and vary the initial spectrum as shown in Figure 9(a). We repeat the estimation process as done in Section 3.1, but now with a wider range of $\lambda \in \{20, 30, \ldots, 2000\}$ as larger regularization parameters are likely to be chosen when the initial spectrum is more precise. Figure 9(b) shows the RMSE plots for the three initial values, each of which are averaged over 20 trials. As clearly indicated by the plot, spectrum estimation with better initial guess leads to improvements in the accuracy and appears to be more stable against $\lambda$, compared to the poor initial guess. This finding is further confirmed by comparing Figure 9(c) and (d), where using better initial guess, Figure 9(c), results in narrower band between the spectral curves than the poor initial guess, Figure 9(d).

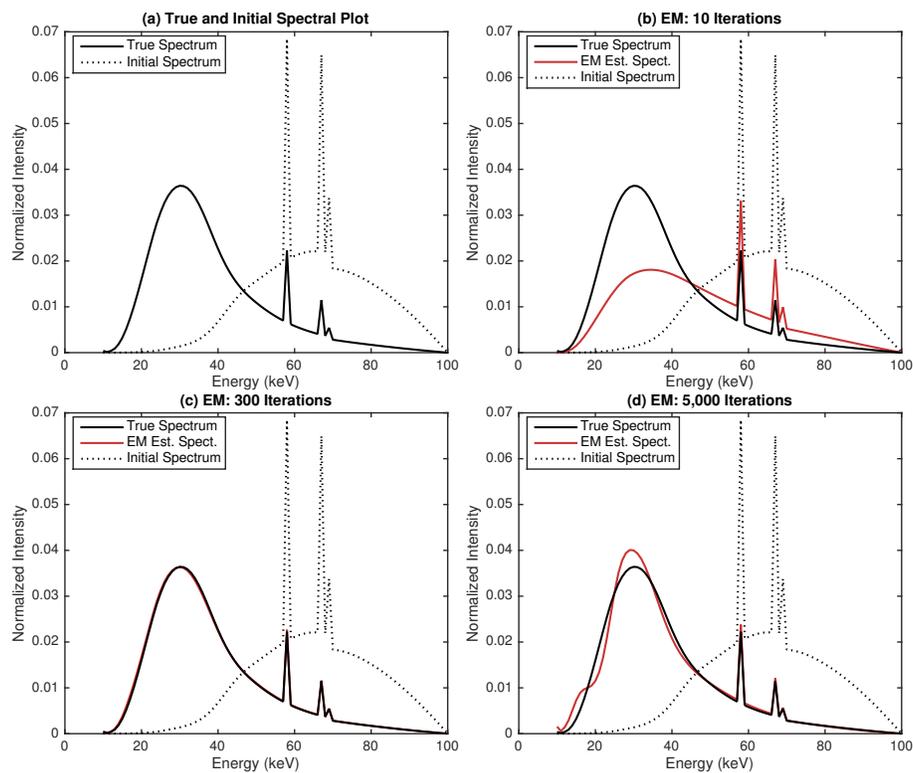

Figure 8: Spectrum estimation from simulated transmission measurements by use of EM. A detailed description of the simulation setting is given in the simulation study (Section 3.1). Panel (a) shows the ground truth x-ray spectrum (black solid line) and the initial x-ray spectrum (black dotted line). The remaining three panels show the estimated spectra by EM for different number of iterations.

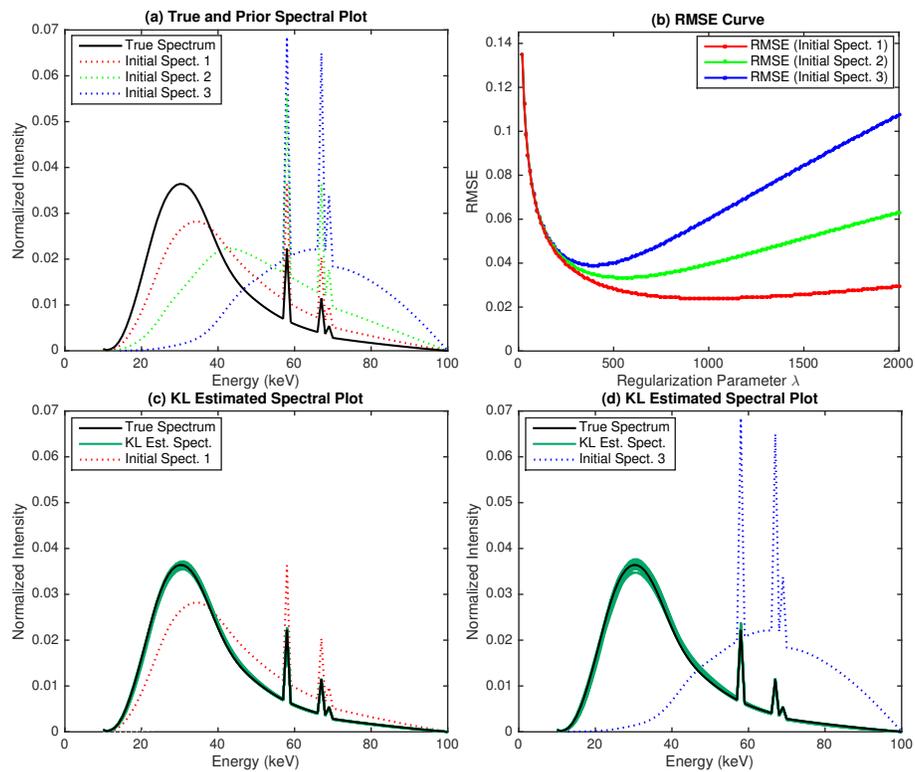

Figure 9: Spectrum estimation from simulated transmission measurements by use of KL. Different types of true and initial x-ray spectrum are employed as depicted in panel (a). In each setting, spectrum reconstruction is performed for 20 independent sets of transmission measurements. Panel (b): Spectral curves for 20 different trials. The band formed by the curves shows variation between the reconstructed x-ray spectra. Panels (c),(d): The RMSE curves computed by (9) for different regularization parameters. Each point represents an average over 20 trials.